\documentclass[12pt]{article}
\usepackage[dvips]{graphicx}
\usepackage{latexsym}

\newcommand{\rad}	{{\rm rad}}
\newcommand{\external}	{{\rm ext}}
\newcommand{\internal}	{{\rm int}}
\newcommand{\trans}	{{\rm trans}}
\newcommand{\sing}	{{\rm sing}}
\newcommand{\Ptan}	{{P_{\tan}}}
\newcommand{\diag}	{{\rm diag}}
\newcommand{\BG}	{{\rm BG}}
\newcommand{\vac}	{{\rm vac}}
\newcommand{\eff}	{{\rm eff}}

\newcommand{\BH}	{{\rm BH}}
\newcommand{\pl}	{{\rm pl}}
\newcommand{\maxi}	{{\rm max}}
\newcommand{\mini}	{{\rm min}}
\newcommand{\peak}	{{\rm peak}}
\newcommand{\rhopeak}	{{\rho\mbox{-}{\rm peak}}}
\newcommand{\Gpeak}	{{G\mbox{-}{\rm peak}}}

\newcommand{\fig}[1]	{Figure \ref{#1}}

\newcommand{\EQ}[1]	{(\ref{#1})}

\begin{document}

\begin{flushright}
 \begin{minipage}[b]{43mm}
  hep-th/0310185\\
  WIS/27/03-OCT-DPP\\
 \end{minipage}
\end{flushright}

\renewcommand{\thefootnote}{\fnsymbol{footnote}}
\begin{center}
 {\Large\bf
 Radiation Ball as a Black Hole
 }\\
 \vspace*{3em}
 {Yukinori Nagatani}\footnote
 {e-mail: yukinori.nagatani@weizmann.ac.il}\\[1.5em]
 {\it Department of Particle Physics,\\
 The Weizmann Institute of Science, Rehovot 76100, Israel}
\end{center}
\vspace*{1em}

\begin{abstract}
 A structure of the radiation-ball
 which is identified as a Schwarzschild black hole
 is found out
 by investigating the backreaction of Hawking radiation into space-time.
 The structure consists of the radiation
 which is gravitationally bounded in the ball
 and of a singularity.
 The entropy of the radiation in the ball
 is proportional to the surface-area of the ball and
 nearly equals to the Bekenstein entropy.
 The Hawking radiation is regarded as a leak-out from the ball.
 %
 %
 There arises no information paradox because 
 there exists no horizon in the structure.
\end{abstract}

%

\newpage
\section{Introduction}\label{intro.sec}

One of the most natural approaches to understand
the quantum-mechanical properties of a black hole,
e.g. the Hawking radiation \cite{Hawking:1975sw,Hawking:1974rv}
and the origin of the Bekenstein entropy \cite{Bekenstein:1973ur},
is considering the structure of the black hole.
The membrane-descriptions \cite{Thorne:1986iy},
the stretched horizon models \cite{Susskind:1993if,Susskind:1994sm}
and the quasi-particles model \cite{Iizuka:2003ad}
were proposed as this kind of approach.
The D-brane descriptions of the (near) extremal-charged black hole
succeeded in deriving the entropy \cite{Strominger:1996sh}
and the Hawking radiation \cite{Horowitz:1996fn}.
These approaches are just considering the dual structure of the black hole.

Hotta proposed the Planck solid ball model \cite{{Hotta:1997yj}}
based on the stretched horizon model.
The structure of the Schwarzschild black hole
consists of a Planck solid ball and
a layer of radiation around the ball.
There is no horizon and 
the radius of the ball is slightly greater
than the Schwarzschild radius.
The Planck solid is a hypothetical matter
which arises by the stringy thermal phase transition
due to the high temperature of the radiation.
The temperature of the radiation around the ball
(near the horizon-scale) becomes very high
$T(r) \propto 1/\sqrt{g_{tt}(r)}$
because of the blue-shift effect by the deep gravitational potential
\cite{Susskind:1994sm,Iizuka:2003ad,Hotta:1997yj,Nagatani:2003ps}.
Both the Bekenstein entropy and the mass of the black hole
are carried by the radiation.

The Planck solid ball is proposed
as a backreaction of the high temperature radiation
into the space-time structure.
If there arises no Planck solid transition,
it is natural to ask what occurs as the backreaction.
In this paper
we consider the backreaction of the thermal radiation
by solving the Einstein equation.
We obtain a structure of
{\it the gravitationally-bounded radiation-ball with a singularity}
as a solution.
The structure is spherically symmetric and
is characterized by a parameter $r_\BH$
which means the radius of the radiation-ball.
The structure has no horizon and
the gravitational potential in the ball
becomes quite deep $g_{tt} \sim 10^{-3}/(m_\pl^2 r_\BH^2)$.
The exterior ($r > r_\BH$) of the ball
is identified as the Schwarzschild black hole with radius $r_\BH$.
In the interior ($r < r_\BH$)
there is a high density radiation $\rho \sim m_\pl^4$
with the Planck temperature $T \sim m_\pl$.
The radiation is trapped in the ball
by the deep gravitational potential.
The total entropy of the radiation
is proportional to the surface-area of the ball
and is near the Bekenstein entropy.
%
The Hawking radiation is regarded as
a leak-out of the radiation from the surface of the ball.
The singularity in the structure is naked
and has quite large negative mass, however,
the total mass of the structure (the radiation and the singularity)
becomes the ordinary mass of the black hole with radius $r_\BH$.
Therefore the Schwarzschild black hole with the Hawking radiation
is identified as our structure
and the Bekenstein entropy is carried by the radiation in the ball.

The plan of the rest of the paper is the following.
In the next section 
we present a setup and why the backreaction should be considered.
In Section \ref{solution.sec},
we derive the solution of the radiation-ball.
In Section \ref{entropy.sec},
we calculate the entropy of the radiation.
In Section \ref{singularity.sec},
we consider the properties of the singularity.
In Section \ref{particle-motion.sec},
we consider the motion of the test particle in the structure.
In the final section we give discussions.

\section{Setup and Backreaction}\label{backreaction.sec}

The generic metric of the spherically symmetric static space-time is 
\begin{eqnarray}
 ds^2
  &=&
  F(r) dt^2 - G(r) dr^2 - r^2 d\theta^2 - r^2 \sin^2\theta \, d\varphi^2,
  \label{generic-metric}
\end{eqnarray}
which is parameterized by the time coordinate $t$
and the polar coordinates $r$, $\theta$ and $\varphi$.
The elements $F(r)$ and $G(r)$ are functions depending only on $r$.

The Hawking temperature of a Schwarzschild black hole
with radius $r_\BH$ is given by
\begin{eqnarray}
  T_\BH &=& \frac{1}{4\pi} \frac{1}{r_\BH}.
\end{eqnarray}
We put the black hole into the background of the radiation
with the temperature $T_\BH$
to consider the stationary situation (or equilibrium) of the system
\cite{Gibbons:1976ue}.
The energy density of the background radiation ($r\rightarrow\infty$)
is given by the thermodynamical relation:
\begin{eqnarray}
 \rho_\BG &=& \frac{\pi^2}{30} g_* T_\BH^4,
  \label{rho-BG}
\end{eqnarray}
where $g_*$ is the degree of the freedom of the radiation.
We assume that $g_*$ is a constant
in order to simplify the analysis.
In the situation
we expect the local temperature distribution of the radiation as
\begin{eqnarray}
 T(r) &:=& \frac{T_\BH}{\sqrt{F(r)}} \label{Local-Temperature}
\end{eqnarray}
due to the effect of the gravitational potential $F(r) = g_{tt}(r)$
\cite{Hotta:1997yj,Nagatani:2003ps}.
Therefore the energy-density and the pressures of the radiation become
\begin{eqnarray}
 \rho(r)  &=&             \frac{\rho_\BG}{F^2(r)} \label{rho} \\
 P_r(r)   &=& \frac{1}{3} \frac{\rho_\BG}{F^2(r)}, \label{Pr} \\
 \Ptan(r) &=& \frac{1}{3} \frac{\rho_\BG}{F^2(r)}, \label{Ptan}
\end{eqnarray}
respectively.
$P_r(r)$ is the pressure in the $r$-direction and
$\Ptan(r)$ is the pressure in the tangential direction
($\theta$- and $\varphi$- direction).

If we fix the space-time structure in the Schwarzschild metric:
\begin{eqnarray}
 F_\BH(r) &=& 1 - \frac{r_\BH}{r}, \label{SSF}\\
 G_\BH(r) &=& \left[1 - \frac{r_\BH}{r}\right]^{-1}, \label{SSG}
\end{eqnarray}
then both the density \EQ{rho} and the pressures (\EQ{Pr} and \EQ{Ptan})
diverge on the horizon $r = r_\BH$.
The total radiation-energy around the black hole
$ \int_{r_1}^{r_2} 4 \pi r^2 dr \rho(r) = 1/(r_1 - r_\BH) + \cdots$
also diverges as $r_1 \rightarrow r_\BH$.
This problem requires that we should consider
the backreaction of the radiation into the space-time structure.

\section{Radiation-Ball Solution}\label{solution.sec}

The space-time structure of the black hole
including the backreaction from the radiation
is computed by solving the Einstein equation
$R_\mu^{\ \nu} -\frac{1}{2} R \delta_\mu^{\ \nu} - \Lambda\delta_\mu^{\ \nu}
= (8\pi/m_\pl^2) T_\mu^{\ \nu}$
with the metric \EQ{generic-metric} and
with the energy-momentum-tensor
$T_{\mu}^{\ \nu}(r)
= \diag\left[\; \rho(r), -P_r(r), -\Ptan (r), -\Ptan(r) \; \right]$
whose elements are given by \EQ{rho}, \EQ{Pr} and \EQ{Ptan}.
We have introduced the positive cosmological constant
$\Lambda = (8\pi/m_\pl^2) \rho_\vac$ to stabilize
the universe from the radiation background.
The background space-time becomes the Einstein static universe 
by choosing $\rho_\vac = \rho_\BG$.
%
%
The Einstein equation becomes three equations:
\begin{eqnarray}
  \frac{-G + G^2 + r G'}{r^2 G^2}
   &=&
   \frac{8 \pi}{m_\pl^2} \left\{\rho(r) + \rho_\BG \right\}, \label{Erho}\\
  \frac{F - F G + r F'}{r^2 F G}
   &=&
   \frac{8 \pi}{m_\pl^2} \left\{P_r(r) - \rho_\BG \right\}, \label{EPr}\\
  \frac{
   - r (F')^2 G  - 2 F^2 G'
   - r F F' G'
   + 2 F G (F' + r F'')
  }{ 4 r F^2 G^2}
   &=&
   \frac{8 \pi}{m_\pl^2} \left\{\Ptan(r) - \rho_\BG \right\}, \label{EPtan}
\end{eqnarray}
where $m_\pl$ is the Planck mass.
%
%
%
We obtain the relation
\begin{eqnarray}
 G(r) &=& \frac{1 - \frac{r}{2} \frac{\rho'(r)}{\rho(r)}}
  {1 + \frac{8\pi}{m_\pl^2} r^2
   \left\{\frac{1}{3} \rho(r) - \rho_\BG \right\} }
  \label{G-rho}
\end{eqnarray}
from the equation \EQ{EPr} with \EQ{Pr}.
By substituting \EQ{G-rho} into the equation \EQ{Erho} with \EQ{rho},
we obtain the differential equation for the energy density $\rho(r)$ as
\begin{eqnarray}
 	&& r \rho \left[
		-24		\rho^2
		\left\{ \textstyle 1 - \frac{\rho_\BG}{\rho} \right\}
		+12	r	\rho	\rho' 
		+	r^2	\rho'^2
		\left\{ \textstyle 1 - 9 \frac{\rho_\BG}{\rho} \right\}
		- 2 	r^2	\rho	\rho''
		\left\{ \textstyle 1 - 3 \frac{\rho_\BG}{\rho} \right\}
	\right] \nonumber\\
	&& \;+\; \frac{3 m_\pl^2}{8\pi}
	\left\{
		- 4 \rho \rho'
		+ 3 r \rho'^2
		- 2 r \rho \rho''
	\right\} \;=\; 0. \label{rhoEQ}
\end{eqnarray}
We also obtain the same differential equation \EQ{rhoEQ}
by substituting \EQ{G-rho} into the equation \EQ{EPtan} with \EQ{Ptan},
therefore,
the Einstein equation in \EQ{Erho}, \EQ{EPr} and \EQ{EPtan}
and the assumption of the energy-momentum tensor in
\EQ{rho}, \EQ{Pr} and \EQ{Ptan} are consistent.
Although the number of the differential equations exceeds
the number of the unknown functions, there exists a solution.
This is a non-trivial feature of the system.

\begin{figure}
 \begin{center}
  \includegraphics[scale=0.8]{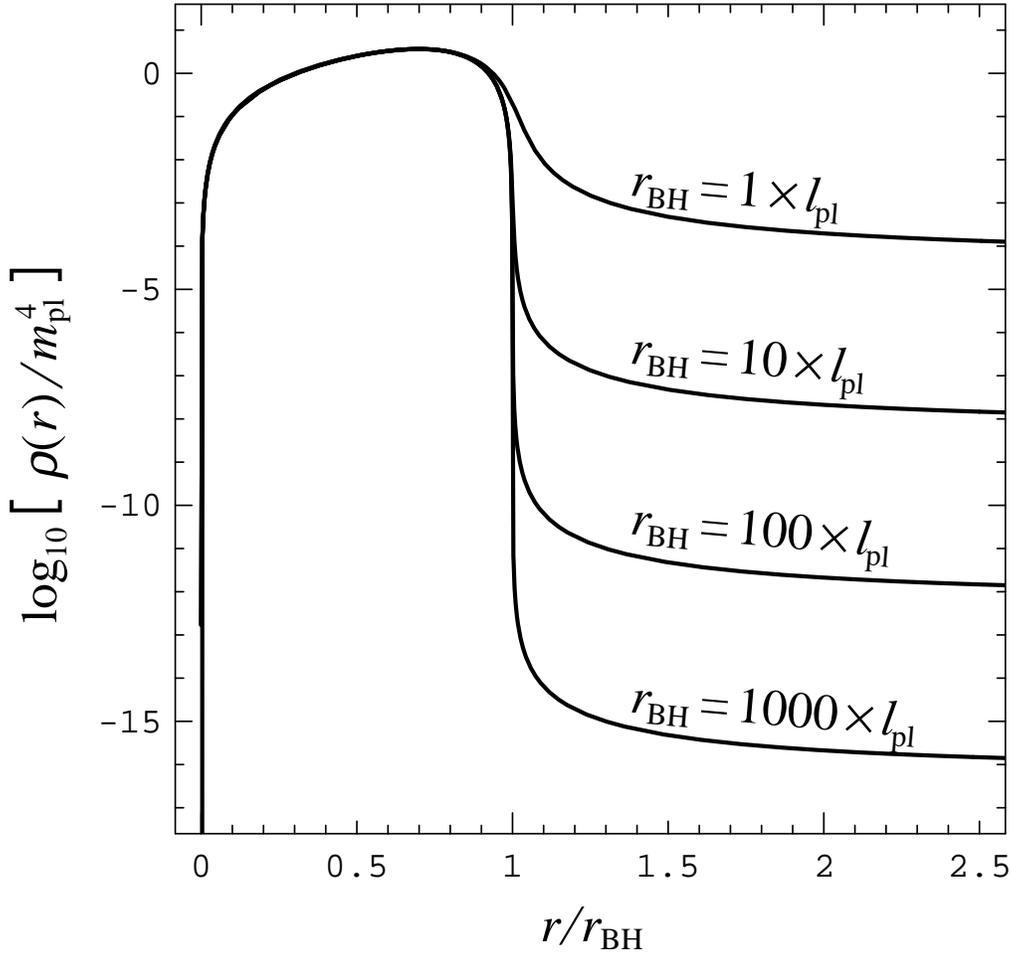}%
 \end{center}
 \caption{%
 Distributions of the radiation-energy-density
 $\rho(r)$ in the solution of the radiation-ball
 for various $r_\BH$.
 The horizontal axis is the coordinate $r$
 normalized by $r_\BH$.
 The solutions for $r_\BH/l_\pl = 1,10,100$ and $1000$ are displayed,
 where $l_\pl := m_\pl^{-1}$ is the Planck length.
 In the figure the degree of the freedom $g_* = 4$ is assumed.
 We find that the distribution in the ball ($r < r_\BH$)
 has a universal form.
 The density $\rho(r)$ approaches to the background density $\rho_\BG$
 defined in \EQ{rho-BG} when $r$ becomes large.
 \label{RHOLogMult.eps}%
 }%
\end{figure}

\begin{figure}
 \begin{center}
  \includegraphics[scale=0.8]{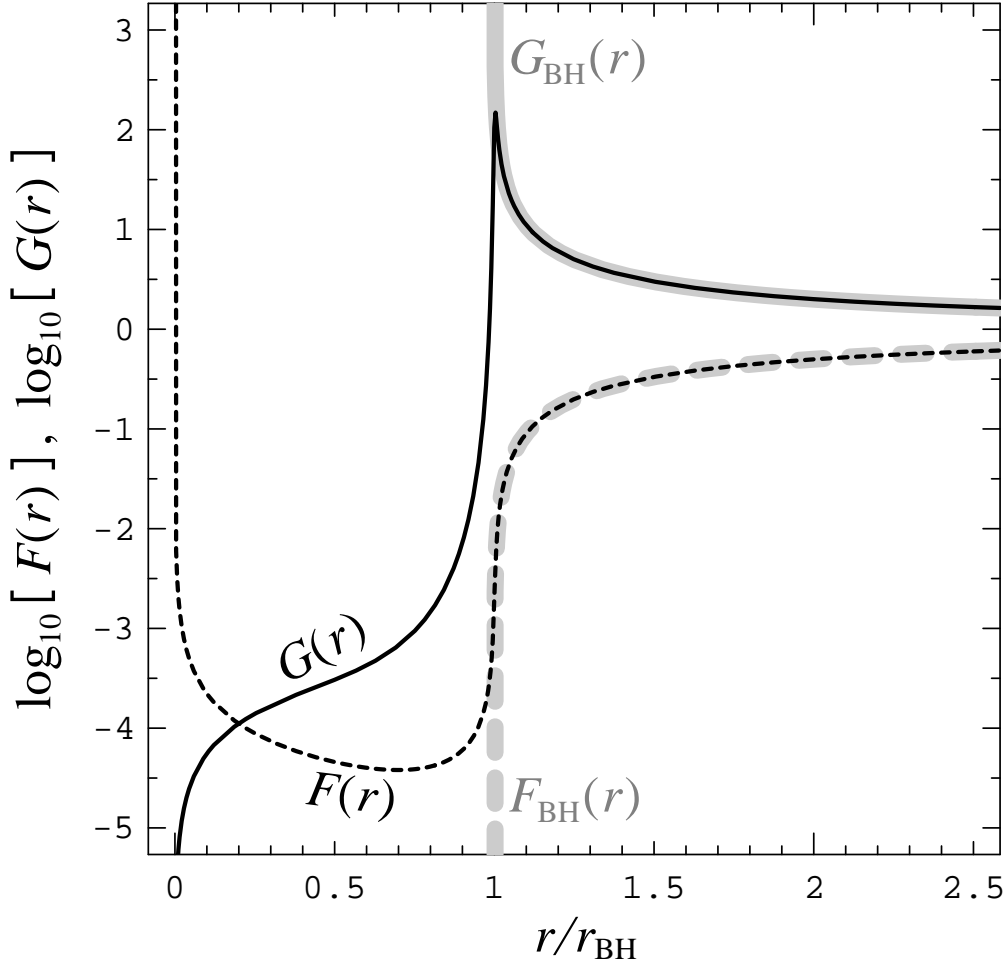}%
 \end{center}
 \caption{%
 Elements of the metric $F(r)$ and $G(r)$ in the solution
 of the radiation-ball
 with $r_\BH = 10 \times l_\pl$ and $g_* = 4$.
 The thin dotted curve is $F(r)$ and
 the thin solid curve is $G(r)$ in the solution.
 The thick dotted gray curve $F_\BH(r)$ and
 the thick solid gray curve $G_\BH(r)$ show
 the external part of the Schwarzschild metric with the radius $r_\BH$.
 Curves for more various parameters are shown in \cite{MathCode:2003}.
 \label{FG.eps}%
 }%
\end{figure}

The differential equation \EQ{rhoEQ} is numerically solved
and we find out the solution
whose exterior part corresponds to
the exterior of the Schwarzschild black hole
with the Hawking radiation in the Einstein static universe.
The Mathematica code for the numerical calculation
can be downloaded on \cite{MathCode:2003}.
The solution is parameterized by a radius $r_\BH$
which is the Schwarzschild radius of the correspondent black hole.
The numerical solutions for various $r_\BH$ are displayed in
\fig{RHOLogMult.eps}.
The element of the metric $F(r)$ is derived by \EQ{rho}
and $G(r)$ is derived by \EQ{G-rho}.
A typical form of the metric elements is displayed in \fig{FG.eps}.
The solution indicates that
most of the radiation is trapped in the sphere
by the gravitational potential $F(r)$ and
the radius of the sphere is given by $r_\BH$.
We call the structure the radiation-ball.

\begin{figure}
 \begin{center}
  \includegraphics[scale=0.8]{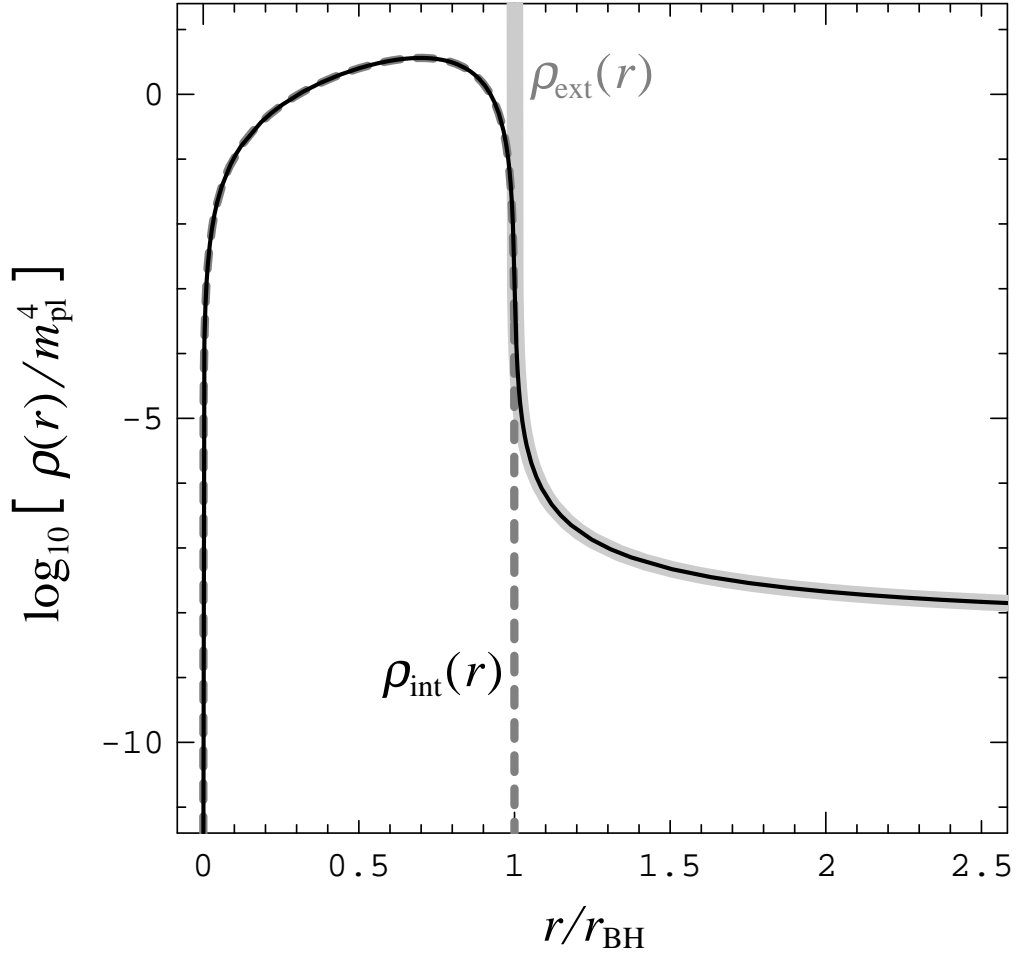}%
 \end{center}
 \caption{%
 A numerical solution $\rho(r)$,
 an approximated solution $\rho_\external(r)$ for the exterior ($r>r_\BH$) and
 an approximated solution $\rho_\internal(r)$ for the interior ($r<r_\BH$).
 The thin solid curve is $\rho(r)$,
 the thick gray curve is $\rho_\external(r)$
 and the dotted gray curve is $\rho_\internal(r)$.
 In the figure we have assumed $r_\BH = 10 \times l_\pl$ and $g_*=4$.
 \label{RHOLog.eps}%
 }%
\end{figure}

For the external region ($r > r_\BH$) the differential equation \EQ{rhoEQ}
is approximated to
\begin{eqnarray}
	- 4 		\rho		\rho'
	+ 3 	r	\rho'^2
	- 2 	r	\rho		\rho''
	&=& 0
\end{eqnarray}
when $r_\BH$ is much greater than the Planck length $l_\pl := m_\pl^{-1}$.
The approximated equation has a solution
\begin{eqnarray}
 \rho_\external(r)
  &=&
  \rho_\BG  \times \left(1 - \frac{r_\BH}{r}\right)^{-2},
\end{eqnarray}
and the elements of the metric become
\begin{eqnarray}
 F_\external(r)
  &=& 1 - \frac{r_\BH}{r}, \label{Fexternal}\\
 G_\external(r)
  &=&
  \left[
   F_\external(r)
     \;+\; \frac{8\pi}{3m_\pl^2} \rho_\BG r^2
           \left\{ F^{-1}_\external(r) - 3 F_\external(r) \right\}
  \right]^{-1}. \label{Gexternal}
\end{eqnarray}
The approximated solution $\rho_\external(r)$
is corresponding to \EQ{rho} with the Schwarzschild metric \EQ{SSF}.
The resultant metric ($F_\external(r)$ and $G_\external(r)$) is also
consistent with the external Schwarzschild metric (\EQ{SSF} and \EQ{SSG})
with a background correction of the Einstein static universe.
%
The function-form of the approximated solution
is displayed in \fig{RHOLog.eps}.

For the internal region ($r < r_\BH$) the differential equation \EQ{rhoEQ}
is approximated to
\begin{eqnarray}
 	-24		\rho^2
	\;+\;12	r	\rho	\rho'
	\;+\;	r^2	\rho'^2
	\;-\;2 	r^2	\rho	\rho''
	&=& 0 \label{RhoOutEQ}
\end{eqnarray}
for $r_\BH \gg l_\pl$.
We obtain the solution of the equation \EQ{RhoOutEQ} as
\begin{eqnarray}
 \rho_\internal(r)
  &=&
  \frac{135}{\pi} \frac{1}{g_*} m_\pl^4
  \left(\frac{r}{r_\BH}\right)^2
  \left[ 1 - \left(\frac{r}{r_\BH}\right)^5 \right]^2,
\end{eqnarray}
and also obtain the elements of the metric
\begin{eqnarray}
 F_\internal(r)
  &=&
  \frac{g_*}{720 \sqrt{2\pi}}
  \frac{1}{m_\pl^2 r_\BH^2}
  \left( \frac{r_\BH}{r} \right)
  \left[1 - \left(\frac{r}{r_\BH}\right)^5\right]^{-1},\\
 G_\internal(r)
  &\simeq&
   \frac{g_*}{72}
   \frac{1}{m_\pl^2 r_\BH^2}
   \left(\frac{r}{r_\BH}\right)
   \left[1 - \left(\frac{r}{r_\BH}\right)^5 \right]^{-3},
\end{eqnarray}
where the coefficients of the solution are determined by
matching to the numerical solution.
The density $\rho_\internal(r)$ has the maximum value
$\rho_\maxi = (125 \cdot 3^{3/5})/(4 \pi \cdot 2^{2/5}) \: m_\pl^4/g_*
\simeq 14.57 \times m_\pl^4/g_*$
on the radius $r_{\rhopeak} = 6^{-1/5} r_\BH \simeq 0.6988 \times r_\BH$.
On the same radius, $F(r)$ has the minimum value
$
F_\mini =
g_*/(200 \cdot 2^{3/10} \cdot 9^{4/5} \sqrt{\pi} m_\pl^2 r_\BH^2)
\simeq 9.515 \times 10^{-4} g_* /(m_\pl^2 r_\BH^2)$.

On the transitional region ($r \sim r_\BH$)
the approximated form is a little complicated:
\begin{eqnarray}
 \rho_\trans(r)
  &=&
 	\frac{3}{8\pi} \frac{m_\pl^2}{r_\BH^2}
	+ \frac{(825)^2}{128 \pi^2}
	\frac{m_\pl^4}{g_*}
	\frac{r - r_\BH}{r_\BH^2}
	\left[
	      \left( r - r_\BH \right)
	    - \sqrt{ \left(r - r_\BH\right)^2
	             + \frac{96\pi}{(825)^2} \frac{g_*}{m_\pl^2}} \:
	\right].
\end{eqnarray}
The element of the metric $G(r)$ has the maximum value
$G_\peak \simeq 30.86 \times m_\pl r_\BH / \sqrt{g_*}$
at the radius $r_{\Gpeak} \simeq r_\BH + 0.01215 \times \sqrt{g_*} l_\pl$
which is quite slightly greater than $r_\BH$ (see \fig{FG.eps}).

\section{Entropy of the Radiation in the Ball}\label{entropy.sec}

The entropy density of the radiation with the temperature $T$
is given by the thermodynamical relation:
\begin{eqnarray}
 s &=& \frac{2\pi^2}{45} g_* T^3.
  \label{entropy-1.eq}
\end{eqnarray}
By combining the form of the energy density
$\rho = \frac{\pi^2}{30} g_* T^4$,
the entropy density is described as a function of the energy density:
\begin{eqnarray}
 s(\rho) &=& \frac{2 \cdot (30)^{3/4} \sqrt{\pi}}{45} g_*^{1/4} \rho^{3/4}.
  \label{entropy-2.eq}
\end{eqnarray}
The total entropy of the radiation
inside ($r < r_\BH$) of the ball becomes
\begin{eqnarray}
 S_\internal
  &\simeq&
  \int_{0}^{r_\BH} 4 \pi r^2 dr \sqrt{G_\internal(r)} \;
  s\left(\rho_\internal(r)\right) \nonumber\\
 &=&
  \frac{(8 \pi)^{3/4}}{\sqrt{5}} m_\pl^2 r_\BH^2
 \;\simeq\; 5.0199 \times \frac{r_\BH^2}{l_\pl^2,}
 \label{entropy-result.eq}
\end{eqnarray}
where $l_\pl := m_\pl^{-1}$ is the Planck length.
The entropy \EQ{entropy-result.eq}
is proportional to the surface-area of the ball,
i.e.,
the area-law of the black-hole-entropy is reproduced.
The entropy \EQ{entropy-result.eq}
is a little greater than the Bekenstein entropy \cite{Bekenstein:1973ur}:
\begin{eqnarray}
 S_{\rm Bekenstein}
  &=& \frac{1}{4} \frac{\rm (Horizon\ Area)}{l_\pl^2}
  \;=\; \pi \times \frac{r_\BH^2}{l_\pl^2,}.
\end{eqnarray}
The ratio becomes $S_\internal/S_{\rm Bekenstein} \simeq 1.5978$.
Therefore the origin of the black hole entropy
is regarded as the entropy of the radiation in the ball.

The information paradox of the black hole does not arise
because the structure of the radiation-ball has no horizon.
The information of the matter which has fallen into the black hole
is carried by the internal radiation of the ball.
The picture that the entropy of the black hole is carried by radiation
is similar to
the quasi particle description in the stretch horizon model
\cite{Iizuka:2003ad} and the Planck solid ball model \cite{Hotta:1997yj}.
In these models the entropy is carried by the particles
around the spherical surface.
On the other hand the particles in the spherical body carries
the entropy in our model.

\section{Singularity}\label{singularity.sec}

The asymptotic solution of the singularity ($r=0$) becomes
\begin{eqnarray}
 F_\sing(r)
  &=&
  \frac{g_*}{720\sqrt{2\pi}}
  \frac{1}{m_\pl^2 r_\BH^2} \left(\frac{r_\BH}{r}\right),\\
 G_\sing(r)
  &=& \frac{g_*}{72 m_\pl^2 r_\BH^2} \left(\frac{r}{r_\BH}\right).
\end{eqnarray}
The mass of the singularity is evaluated as
\begin{eqnarray}
 m_\sing
  &:=&
  \lim_{r \rightarrow 0} \frac{m_\pl^2}{2} r
  \left[1 - \frac{1}{G(r)}\right]
  \ =\  -\frac{36}{g_*} m_\pl^4 r_\BH^3,
\end{eqnarray}
which is negative and
whose absolute value is much greater than the mass of
the correspondent black hole
$m_\BH = (m_\pl^2/2) r_\BH$.
The gravitational mass of the radiation in the ball becomes
\begin{eqnarray}
 m_{\rad}
  &:=&   \int_0^{r_\BH} 4 \pi r^2 dr \rho(r)
  \ \simeq\  \frac{36}{g_*} m_\pl^4 r_\BH^3 \;+\; m_\BH
\end{eqnarray}
which is also much greater than $m_\BH$.
As a consequence of the cancellation of
the large negative singularity-mass $m_\sing$
and the large positive radiation-energy $m_{\rad}$,
the total mass of the structure becomes
the ordinary mass of the black hole
\begin{eqnarray}
 m_\sing \;+\; m_{\rad} &\simeq& m_\BH.
\end{eqnarray}
This property is consistent with
that the external part ($r>r_\BH$) of the structure is essentially
given by the Schwarzschild black hole with mass $m_\BH$.

\section{Particle Motion in the Solution}\label{particle-motion.sec}

%

The radial motion of a test particles in the solution
is described by the geodesic equation
\begin{eqnarray}
 \dot{r}^2(t) &=& -W_\eff\left(r(t)\right),\label{particle-eom-r-2}
\end{eqnarray}
for the particle trajectory $(t, r(t))$,
where we have defined the effective potential
\begin{eqnarray}
 W_\eff(r) &:=&
  - \frac{F(r)}{G(r)} \:
  \left[
    1
    \;-\; \left(\frac{m}{E}\right)^2 F(r)
  \right].
  \label{Particle-Effective-Potential-1}
\end{eqnarray}
The parameter $E$ is the energy of the particle and 
the parameter $m$ is the mass of the particle.
There is repulsive force in the region $0< r < r_\rhopeak$
and attractive force in the region $r_\rhopeak < r$.
If there is enough interaction among the particles
to thermalize the radiation in the ball,
we expect that
the thermodynamics in the gravitational potential $F(r)$
reproduces the temperature distribution \EQ{Local-Temperature}
which we have assumed.
This treatment of the radiation is a kind of the mean-field approximation
and the derived solution is self-consistent.
Most of the particles are trapped in the ball
and the leak-out from the ball is regarded as the Hawking radiation.

\section{Conclusion and Discussion}\label{discussion.sec}

The structure of the radiation-ball
which we have derived by the consideration
of the backreaction of the the Hawking radiation
is identified as the Schwarzschild black hole.
The origin of the Hawking radiation
is explained as a leak-out of the radiation from the ball.
There arises no information paradox
because the structure has no horizon and
the the inside radiation of the ball carries the entropy.

However the structure has several strange features.
There arises a naked time-like singularity.
The mass of the singularity $m_\sing \sim -m_\pl^4 r_\BH^3$
is negative and its absolute value 
is much larger than the mass of the black hole $m_\BH \sim m_\pl^2 r_\BH$.
On the other hand
the total mass of the radiation is positive and
is also much larger than $m_\BH$.
By the cancelation of the large masses,
the total mass becomes the ordinary mass $m_\BH$.
When the mass of the black hole increases,
it seems that the energy of the singularity is transfered
into the energy of the radiation.
We expect these properties of the singularity is explained
by quantum mechanical treatments of the singularity.
The string theory or quantum gravity may be required
to understand the properties
because the temperature of the radiation around the singularity
becomes the Planck energy scale ($\sim m_\pl$).

The entropy of the radiation in the ball is about $1.6$ times
of the Bekenstein entropy.
The quantum entanglement of the particles in the radiation
can reduce the radiation-entropy,
therefore,
the quantum mechanical treatment of the radiation may be effective
and the relations of the entropy in \EQ{entropy-1.eq}
will be deformed.

Phase transitions arise in (or around) the radiation-ball
because the temperature of the radiation in the ball is very high 
\cite{Nagatani:2003ps,Cline:1996mk,Nagatani:1998gv,Nagatani:2001nz,Nagatani:2003pr}.
The sphaleron process is not suppressed in the ball
because the Higgs vev becomes zero,
then the radiation-ball does not conserve the baryon number.
The hyper charged radiation-ball produces the net baryon number
by the effect of the baryon chemical potential \cite{CKN}.
The phase transitions are accompanied
by the formation of the spherical domain walls
whose radii nearly equal to (or are greater than) $r_\BH$.
The generation of the baryon number
\cite{Nagatani:1998gv,Nagatani:2001nz}
and the spontaneous charge-up of the black hole \cite{Nagatani:2003pr}
are expected by the effect of the formed wall.
In this paper
we have assumed the degree of the freedom $g_*$ as a constant
for simplicity.
We should relax this assumption to consider the phase transitions
because the transitions are accompanied by the change of $g_*$.

\begin{flushleft}
 {\Large\bf ACKNOWLEDGMENTS}
\end{flushleft}

 I would like to thank
 Ofer~Aharony, Micha~Berkooz,
 Nadav~Drukker, Bartomeu~Fiol, Hikaru~Kawai, Barak~Kol, Joan~Simon and
 Leonard~Susskind
 for useful discussions.
 I am particularly grateful 
 to Peter Fischer for pointing out a mistake in an earlier version 
 of this paper.
 I would also like to thank the ITP at Stanford university
 and the organizers of the Stanford-Weizmann workshop
 for their hospitality at the early stage of the project.
 I am grateful to Kei~Shigetomi
 for helpful advice and also for careful reading of the manuscript.
 The work has been supported by
 the Koshland Postdoctoral Fellowship of the Weizmann Institute of Science.


\end{document}